# Temperature Dependent Magnetic and Structural Properties of Al Substituted Nanostructured Ferrites with Large Coercive Fields

Pierfrancesco Maltoni[1,2,3]*, Ravi K. Dokala[3], Prativa Pramanik[3], Rafael Araujo[3], Tomas Edvinsson[3], Sergey A. Ivanov[3,4], Bjarne Almqvist[5], Gaspare Varvaro[2], Aldo Capobianchi[2], Nader Yaacoub[6], Charles Hervoches[7], Alberto Martinelli[8], Robert C. Pullar[9], Davide Peddis[1,2]*, Roland Mathieu[3]*

[1]*Department of Chemistry and Industrial Chemistry & INSTM RU, nM2-Lab, University of Genoa, Genoa 16146, Italy*

[2]*Institute of Structure of Matter (ISM), nM2-Lab, Centre for National Research (CNR), Monterotondo Scalo, Rome 00015, Italy*

[3]*Department of Materials Science and Engineering, Uppsala University, Box 35, Uppsala 751 03, Sweden*

[4]*Department of Chemistry, Lomonosov Moscow State University, Leninskie Gory, GSP-1, Moscow, 119991, Russia*

[5]*Department of Earth Sciences, Uppsala University, Villavägen 16, 752 36 Uppsala, Sweden*

[6]*Institut des Molécules et Materiaux du Mans, CNRS UMR-6283, Le Mans Université, F-72085 Le Mans, France*

[7]*Nuclear Physics Institute of the CAS, Řež, Czech Republic*

[8]*CNR-SPIN, Corso Perrone 24, I-16152 Genova, Italy*

[9]*Dipartimento di Scienze Molecolari e Nanosistemi (DSMN), Università Ca' Foscari Venezia, Venezia Mestre, Venezia, 30172, VE, Italy*

Keywords: Chemical Structure, Magnetic Properties, Neutron Diffraction, Raman Spectroscopy, Nanoparticles

*Corresponding authors

**Abstract**

We report a comprehensive study of the temperature-dependent structural, magnetic, vibrational, and dielectric properties of Al-substituted M-type hexaferrites $SrFe_{12-x}Al_xO_{19}$. Neutron powder diffraction and Mössbauer spectrometry show that $Al^{3+}$ preferentially replaces $Fe^{3+}$ at spin-up octahedral sites ($2a$, $12k$), disrupting the exchange coupling with the spin-down $4f$ tetrahedral sites and leading to a progressive reduction of site-specific magnetic moments and a systematic decrease in the Curie temperature, supported by temperature-dependent susceptibility measurements. Raman spectroscopy reveals pronounced phonon anomalies near $T_C$, particularly in modes associated with bipyramidal Fe–O vibrations, reflecting the weakening of both $4e$–$12k$ and $4e$–$4f$ exchange pathways. However, the coercive field exhibits a dramatic increase, reaching $\mu_0 H_C \sim 1.2$ T for $SrFe_{9.6}Al_{2.4}O_{19}$, among the largest values reported for this class. Susceptibility measurements suggest that Al substitution, while weakening the superexchange network, contributes to the stabilization of single-domain behavior.

**Introduction**

Permanent magnets (PMs) are fundamental to a wide range of energy technologies, including electric motors, generators, wind turbines, and magnetocaloric refrigeration systems [1–3]. The global push toward sustainable energy solutions continues to fuel the demand for cost-effective and environmentally friendly magnetic materials [4,5]. Among these, hexaferrites, already widely used in commercial permanent magnets, such as $SrFe_{12}O_{19}$ remain highly attractive not only for their high coercivity, chemical stability, abundance and low cost, but also because tailoring their properties to enhance the $(BH)_{max}$ could open new opportunities to employ them as substitutes for NdFeB magnets in applications where extremely high performance is unnecessary and rare-earth solutions are excessively costly [6–8].

The crystal structure of M-type hexaferrites offers multiple opportunities to modulate their magnetic properties through targeted substitution at specific cationic sites within the magnetoplumbite structure. Each of the five crystallographically distinct cationic sites ($2a$, $4e$, two $4f$, and $12k$) contributes differently to the total magnetization and magnetic anisotropy due to their individual spin alignments and local coordination environments [9]. The seminal review by Kojima[10] laid the foundation for understanding these substitution mechanisms, although comprehensive reviews were only recently published with the works of R. C. Pullar[11], S. Mahmood[12], and D. Lisjak et al.[13]. The primary objective of cation substitution is to tune the intrinsic magnetic interactions governed by superexchange pathways, whose strength and nature are highly sensitive to cation type, charge state, and site occupancy [14,15]. Importantly, the site-specific contribution to magnetic properties is also influenced by the spin state and orbital moment of the substituted ion, which together determine both net magnetization and magnetic anisotropy. One established strategy to enhance the performance of ferrite magnets involves partial substitution of $Fe^{3+}$ with diamagnetic cations, such as $Al^{3+}$. This doping has been shown to increase coercivity significantly without introducing secondary impurity phases [11,16,17]. This effect is attributed to the preferential occupation of Fe sites with spin-up contributions in the ferrimagnetic lattice, leading to a reduction in saturation magnetization but a concurrent increase in magnetic coercivity [18] (here and in what follows, 'spin-up' and 'spin-down' refer not to absolute configurations, but to the relative orientation of moments in two opposing sublattices). While the structural and magnetic properties of these materials at 300 K have been extensively studied[19,20], their temperature-dependent behaviour remains less explored, particularly in view of high-temperature applications[21–23]. The Curie temperature ($T_C$) typically lies in the range of 720–740 K for pure $BaFe_{12}O_{19}$ and $SrFe_{12}O_{19}$, and arises from strong superexchange interactions between $Fe^{3+}$ ions mediated by $O^{2-}$ across five magnetically active sublattices[24]. Substitution of $Fe^{3+}$ by non-magnetic cations such as $Al^{3+}$ or $Ti^{4+}$ systematically reduces $T_C$ due to disruption of the Fe–O–Fe superexchange network. For example, in $SrFe_{12-x}Al_xO_{19}$, the $T_C$ decreases by more than 100 °C for x ~ 2, highlighting the destabilizing effect of diamagnetic doping[18]. Similarly, $Ti^{4+}$ substitution leads to a drop in $T_C$ not only by replacing magnetic centers but also via charge compensation mechanisms, which may result in the formation of $Fe^{2+}$, further weakening exchange interactions[25]. Magnetic dopants such as $Co^{2+}$ or $Cr^{3+}$ tend to have a milder impact, though they may still reduce $T_C$ depending on their spin configuration and site occupancy[26]. $Co^{2+}$, for instance, shows weaker Co–O–Fe interactions, marginally reducing $T_C$, though less significantly than $Al^{3+}$ [27,28]. In contrast, $Cr^{3+}$ can partially preserve the magnetic ordering temperature due to its similar ionic radius and spin state[16,29]. Substitution at the A-site ($Ba^{2+}$/$Sr^{2+}$) also influences $T_C$ indirectly through lattice contraction or distortion[30]. $Sr^{2+}$ substitution often leads to a slight increase in $T_C$ compared to $Ba^{2+}$, attributed to stronger crystal fields and smaller ionic radius, which enhance the Fe–O–Fe bond strength. This temperature-induced softening/hardening must be considered when employing doped hexaferrites in high-temperature environments, where optimization of magnetic hardness (via doping) must be balanced against thermal robustness. Yet, key physical properties such as magnetic anisotropy, spin ordering, and dielectric response are inherently temperature-dependent and must be fully characterized to evaluate the functional reliability of these materials under operating conditions. Furthermore, the preparation of ferrites with a modified stoichiometry (and in turn tunable magnetic properties) represents a strategic step also in the design of magnetic nanocomposites[18,31,32]. The use of such modified ferrites as building blocks in nanocomposite systems (e.g., hard–soft magnetic heterostructures) opens up opportunities for engineering hybrid materials in which both microstructure and interphase magnetic interactions can be precisely controlled to enhance overall performance.

In this paper, we present a comprehensive temperature-dependent study of $SrFe_{12-x}Al_xO_{19}$ ferrites, focusing on a selected series with $1 \leq x \leq 2.4$. The evolution of the structural, vibrational, magnetic, and dielectric properties has been investigated from low to high temperatures using a multi-technique approach (i.e., neutron powder diffraction, Raman and Mössbauer spectrometry, magnetic susceptibility, and dielectric measurements). Neutron powder diffraction enables an accurate determination of the cationic distribution and provides direct insight into the temperature evolution of site occupancies, lattice parameters, and local structural distortions. Raman spectroscopy provides insights into spin–phonon coupling and lattice vibrations. The temperature-dependent magnetization probes magnetic transitions and spin reorientation process, while dielectric measurements allow us to monitor the effect of Al on the carrier hopping along Fe-O-Fe bonds.

By correlating these temperature-dependent observables with Al content, we gain a comprehensive understanding of the thermal stability and functional tunability in Al-substituted hexaferrites, thereby providing design guidelines for their implementation in devices operating over broad temperature ranges.

**Methods**

SrFe$_{12-x}$Al$_x$O$_{19}$ (x = 1, 1.4, 2, 2.4) and SrFe$_{12}$O$_{19}$ (were prepared by sol–gel combustion from Fe(NO$_3$)$_3$·9H$_2$O, Al(NO$_3$)$_3$·9H$_2$O, and Sr(NO$_3$)$_2$, with (Fe$^{3+}$+Al$^{3+}$)/Sr$^{2+}$ = 11. After gelation (citric acid, pH=7), the dried gel was combusted at 300 °C, ground, and annealed at 1100 °C for 6 h in air. This series is referred to as SFAO_x and SFO, respectively. More info on the synthesis is reported in Ref.[18].

The powder samples (~2.5 g) were characterized by neutron powder diffraction (NPD) at 15, 298 (room temperature), 500 and 800 K on the instrument MEREDIT at the Nuclear Physics Institute (NPI), ASCR, Rez, Czech Republic (http://neutron.ujf.cas.cz/meredit). During the measurements the samples were placed in the vanadium container with diameter of 10 mm. The height of the powder in the container was about 20 mm. Sample slits were selected to follow the size of the powder in the container (~12 × 20 mm). Room temperature measurements were performed using the carousel-automatic sample exchanger and low temperature measurements were performed using close-cycle cryostat from Scientific Products Ltd. Temperature stabilized at 10 ± 0.2 K. A mosaic copper monochromator on reflection (220) which produces neutrons with wavelength of 1.46 Å was used. Data were collected between 4° and 144° 2θ with steps of 0.08°.

Structural and magnetic refinements were performed using the Rietveld method[33], via the program FullProf (see SI for more details on the refinement)[34]. The diffraction peaks were described by a modified Thompson–Cox–Hastings pseudo-Voigt function. For the Rietveld analysis, the zero shift, unit cell parameters, atomic positions and isotropic atomic displacement parameters were refined, as well as the Lorentzian refinable size (Y) and strain (X) parameters. In all cases, a powder diffraction pattern of a NIST LaB$_6$ 660b standard was collected for corrections.

Magnetic mass susceptibility ($\chi_{mass}$) was collected as a function of temperature with an MFK1-FA susceptibility bridge (www.agico.cz).

$^{57}$Fe Mössbauer spectra were recorded using a $^{57}$Co/Rh γ-ray source mounted on an electromagnetic transducer with velocity modulated according to a triangular waveform. The spectra were obtained at 300 without an external applied field. The hyperfine structure was modelled by means of a least-square fitting procedure involving Zeeman sextets composed of Lorentzian lines. The isomer shift (IS) values were referred to that of α-Fe at 300 K.

Raman spectroscopy was measured using a Renishaw InVia Reflex spectrometer coupled with 1024-pixel Rencam CCD detector, 2400 lines per millimeter grating, and a 20x objective to focus both the incident laser beam into a spot with diameter of 4–5 μm. The Raman shift was recorded with a frequency-doubled Nd:YAG laser (532 nm), chosen to minimize absorption issues commonly encountered with iron compounds when using red lasers, in the wave number range 100 to 900 cm$^{-1}$ using a spectral resolution of 2 cm$^{-1}$ per detection pixel. The scanning parameter for each Raman spectrum was fixed to 1 s exposure time, and 100 scans were accumulated for each experimental run[35]. Multiple scans on different areas of the same sample were performed with lower laser power to ensure spectral reproducibility. Temperature dependent (300 to 800 K) measurements were performed using a THMS600 stage from Linkam Scientific Instruments. All experiments were carried out using incident powers of about 5mW after confirming no damaging to sample from local heating using a Thorlabs PM160T power meter. Each batch of measurements were calibrated with a standard Si reference line of 520.5 cm$^{-1}$.

First-principles calculations were carried out within the framework of density functional theory (DFT) using the Projector Augmented Wave (PAW) formalism as implemented in the Vienna Ab initio Simulation Package (VASP)[36,37]. Exchange–correlation effects were calculated using the Perdew–Burke–Ernzerhof (PBE) functional within the generalized gradient approximation (GGA)[38]. To properly capture the on-site electronic correlations associated with the Fe 3d states, we adopted the rotationally averaged DFT+U formulation of Dudarev et al.[39], which introduces an effective Hubbard interaction $U_{eff}$ =U-J. A value of $U_{eff}$ =5.3 eV was assigned to the Fe 3d manifold, as this parameter yields a band gap consistent with experimental observations[40]. The most stable magnetic ordering option presents the spin of the Fe ions located in $4f_1$ and $4f_2$ sites antiparallel with that of Fe ions in other sublattices and the net magnetic moment of the cell is 40$\mu_B$. The plane-wave basis was truncated at an energy of 520 eV and Brillouin-zone integrations were performed using a 3×3×1 Monkhorst–Pack grid. Structural relaxations were carried out until the residual forces on all atoms were below $10^{-3}$ eV/Å and the electronic self-consistency loop was converged to below $10^{-8}$ eV. Phonon frequencies and eigenmodes at the Γ point were obtained via density functional perturbation theory (DFPT). A detailed description of the computational workflow, including the derivation of Raman intensities, is provided in the Supporting Information.

Dielectric properties were measured in two electrode configurations with an LCR meter (Model: Agilent E4980A) in a wide range of frequencies (100 Hz - 2 MHz) and temperatures (300 - 800 K). The two-electrode configuration was prepared after polishing and coating each surface of a dense pellet (prepared using a hydraulic press, with a pressure of approximately 2.5 GPa, and sintered for 1h at 1100 °C) with silver paint. The temperature dependent studies were performed using a cryofurnace from Linkam Scientific Instruments (HFS600E-PB4 stage and TMS93 temperature controller) in non-reducing (air) atmosphere. The capacitance (C) and the loss-tangent (tan δ) of sample were recorded as a function of temperature while heating at a rate of 10 K/min. The real part of the relative dielectric constant, $\varepsilon_r'$ was determined from the measured C value according to $\varepsilon_r' = Cd/\varepsilon_0 A$, where $\varepsilon_0$, A, and d, represent the dielectric vacuum permittivity, area and the thickness of the pellet, respectively.

**Results and Discussion**

*Crystal and Magnetic Structure of Al-Substituted Sr-Hexaferrites*

Neutron powder diffraction (NPD) was employed to determine the crystallographic and magnetic structures of the $SrFe_{12}O_{19}$ (SFO) and $SrFe_{12-x}Al_xO_{19}$ (SFAO_x, with x:1, 1.4, 2, 2.4) series at 298 K. The magnetoplumbite structure ($P6_3/mmc$) was confirmed across the series by Rietveld refinement (figure S1 shows Rietveld refinement plots obtained using XRPD data data)[18]. In the refinements, the total Al content was initially assigned uniformly across all Fe sites, after which the relative occupancies were allowed to vary freely so that the $Al^{3+}$ distribution could adjust according to the diffraction data. For reference, the refined numerical occupancies, atomic coordinates, isotropic displacement parameters, for the x = 0, 1, 2 samples are summarized in Table S3.

The refined lattice parameters decrease systematically with increasing $Al^{3+}$ content, consistent with the smaller ionic radius of $Al^{3+}$ compared to $Fe^{3+}$ (0.535 Å and 0.645 Å in octahedral coordination, respectively). For x = 1, a and c are reduced by ~0.3 %, relative to unsubstituted SFO, with further contraction at higher substitution levels (by ~0.9 % for x = 2.4). The Sr and

O atomic positions were refined with full occupancy and constrained thermal parameters, while Fe and Al occupancies were refined freely within each crystallographic site. Maintaining charge neutrality is a critical consideration in substitutional doping of hexaferrites. When $Fe^{3+}$ is substituted with divalent (e.g., $Zn^{2+}$, $Co^{2+}$) or tetravalent ions (e.g., $Ti^{4+}$, $Zr^{4+}$), charge compensation mechanisms must occur to balance the ionic disparity, often resulting in the partial reduction of $Fe^{3+}$ to $Fe^{2+}$. As a result, the magnetic exchange interactions are modified and unwanted electrical conductivity can be introduced, which may be detrimental for applications requiring high-resistivity ferrites[11]. In contrast, the isovalent substitution of $Al^{3+}$ for $Fe^{3+}$ preserves charge balance without the formation of compensating defects or redox-active species. As a result, the structural and magnetic changes observed here are driven primarily by ionic size mismatch and site preference, rather than by valence-induced charge disorder[41].

Site occupancy refinements reveal that $Al^{3+}$ preferentially substitutes $Fe^{3+}$ in the $2a$ and $12k$ octahedral sites, with a smaller but significant fraction entering the $4f_2$ (octahedral) site. Only a minor fraction (< 5 %) is found in the $4f_1$ (tetrahedral), while higher substitution of $4e$ (trigonal bipyramidal) site was found for $x >2$. This selective substitution is consistent with crystal-field stabilization considerations and previous reports on $Al^{3+}$ incorporation into M-type hexaferrites[16,42–44]. With increasing $x$, the relative fraction of $Al^{3+}$ in the 2a site rises sharply (22 % for $x = 1$ to 54 % for $x = 2.4$), while the $12k$ site maintains a steady occupancy of ~12–28 %[18]. Theoretical calculations and experimental findings converge in identifying the preferential occupancy of $Al^{3+}$ at the $2a$ and $12k$ sites[18,45,46], although partial occupation of spin-down sites such as octahedral $4f$ has also been reported[47].

For SFO, SFAO_1 and SFAO_2 samples the magnetic structure was refined from NPD data (**Figure 1** shows the case of SFAO_1). The magnetic reflections were indexed with a propagation vector **k** = (0, 0, 0) and were found to be consistent with a collinear ferrimagnetic arrangement in which the $Fe^{3+}$ moments are aligned parallel or antiparallel to the $c$ axis, depending on the site symmetry (see Supplementary for the detailed representation analysis). This configuration agrees with the uniaxial magnetocrystalline anisotropy characteristic of the M-type hexaferrite structure. The absence of additional magnetic reflections violating the $P6_3/mm'c'$ magnetic symmetry further confirms the purely uniaxial nature of the magnetic ordering. See the supporting information and Tables S1,2 for details of the refinement parameters and site-specific magnetic moments. In all cases, the refinements yielded low $R$ factors (Table S3), indicating an excellent fit between experimental and calculated profiles. No secondary phases were observed within the detection limits, even for the samples with higher $Al^{3+}$ contents. **Figure 2a** shows the refined magnetic moments of $Fe^{3+}$ ions at each crystallographic site for SFO, SFAO_1 and SFAO_2 at 298 K; (see also the representation of the crystal and magnetic structure **Figure 2b**). In unsubstituted SFO, the trigonal bipyramidal $4e$, tetrahedral $4f$ and octahedral $2a$ sites carry the largest moments (~4.26, ~3.93 and ~3.90 $\mu_B$, respectively), whereas the octahedral $4f$ and $12k$ sites exhibit reduced magnitudes. Substitution with $Al^{3+}$ causes a marked decrease in the magnetic moments, particularly $12k$, $2a$ and $4e$ (~20% for SFAO_2), directly reducing the net magnetization. The tetrahedral $4f$ site is less affected, while the octahedral $4f$ shows small variations within experimental uncertainty.

The saturation magnetization $M_s$ measured at 300 K (**Figure 2c**) decreases from ~75 A m$^2$/kg for SFO to ~48 A m$^2$/kg for SFAO_2, in excellent agreement with the net magnetization values ($M_{NPD}$) obtained by summing the refined site moments, according to $M_{NPD}$ [A m$^2$/kg] = $N_A\mu_B$

$\mu_{tot}/M_{weight}$, where $N_A$ is the Avogadro constant, $\mu_B$ the Bohr magneton, $M_{weight}$ the molecular weight of the material, and $\mu_{tot}$ the net magnetic moment per formula unit ($\mu_{tot}(T)$ [$\mu_B$/f.u.] = $6m_{12k}(T) + 1m_{2a}(T) + 1m_{2b}(T) - 2m_{4f1}(T) - 2m_{4f2}(T)$) at the temperature T, and $m$ is the site-averaged magnetic moments per Fe ion. This confirms that the reduction in $M_s$ is driven by site-selective dilution of the spin lattice rather than spin canting or structural disorder. Importantly, the 2a and 12k sites correspond to parallel spin-up sublattices in the collinear ferrimagnetic structure of M-type hexaferrites, so their progressive dilution directly weakens the net magnetic moment. The site-selective replacement of $Fe^{3+}$ mainly in the 2a and then 12k sites explains the pronounced magnetization drop: the total moment decreases by ~32 % from SFO to SFAO_1, in line with the increase of $Al^{3+}$ substitution at the 2a and 12k sites, and by a further ~50 % in SFAO_2 due to the larger occupation of 2a-, 12k-, and, at high $x$, even bipyramidal 4e-sites.

Comparison of our refined site occupancies with literature values for Al-substituted SFO shows similar preferential filling of the 2a and 12k sites, although in our samples the 12k substitution fraction increases more gradually with Al content [19,46]. This may reflect differences in synthesis conditions or $Al^{3+}$ distribution equilibria during high-temperature annealing[46]. The trend in site-resolved magnetic moments (**Figure 2a**) indicates a steady decrease at the 12k site and at the 2a site, confirming that the dominant mechanism for magnetization loss is direct replacement of $Fe^{3+}$ in the main spin-up sublattice, in agreement with magnetometry across the series (Figure S2). The ratio of remanent magnetization to saturation magnetization ($M_R/M_S$) remains nearly constant at ~0.5 for all samples, confirming that the substitution does not alter the uniaxial character of the magnetocrystalline anisotropy[11]. Interestingly, when 4e (bipyramidal) is not substituted, its refined $Fe^{3+}$ moment decreases markedly. This is expected from the topology of exchange paths: the 4e - $Fe^{3+}$ has few Fe–O–Fe links and sits in the Sr layer, so its moment is stabilized indirectly through superexchange with neighboring 12k and 4f sites (**Figure 2d**). Depleting 12k (and partly 4f) with nonmagnetic Al weakens those paths, so 4e's moment softens (even without Al on 4e itself). Concomitantly, the unit-cell contraction with Al (altered distances and bond angles, see Table S4) further reduces superexchange strength, reinforcing the 4e drop. The same exchange-weakening explains the monotonic decrease of $T_C$ with Al[18]: less magnetic Fe on key exchange nodes (2a, 12k) turns into weaker net exchange, and thus lower $T_C$. The resulting modification of local crystal fields and magnetic sublattice interactions contributes to enhanced coercivity (see Figure S2), albeit at the expense of a reduced Curie temperature due to the weakened $Fe^{3+}$–O–$Fe^{3+}$ superexchange[48,49].

*Local Magnetic Environment*

To complement the long-range structural and magnetic information obtained from diffraction, $^{57}Fe$ Mössbauer spectrometry was employed to determine the cationic distribution and local hyperfine structure (Figure S3). This technique provides a powerful local probe of the iron sites, allowing the extraction of hyperfine parameters that reflect their electronic and magnetic surroundings. The experimental spectra were well reproduced by the fitted components, confirming an excellent agreement between calculated and experimental lines and excluding the presence of secondary Fe-based phases even at the highest Al contents. The evolution of hyperfine parameters with increasing $Al^{3+}$ content is summarized in Table S5. No significant change in isomer shift was detected, confirming the preservation of the $Fe^{3+}$ valence state across all samples. In the spectrum of the parent SFO, the subspectra associated with the 12k and 4e sites are well resolved, allowing a clear tracking of their evolution upon Al substitution.

As the Al content increases, the relative area of the 12$k$ component decreases, accompanied by a gradual reduction of its hyperfine field ($B_{hyp}$ from 41.4 to 39.8 T). Conversely, the 2$b$ contribution increases in intensity, with only a modest decline in $B_{hyp}$ (from 41.0 to 39.4 T), suggesting a redistribution of $Fe^{3+}$ population among adjacent sites. A more pronounced change is observed for the 4$f$ (tetrahedral) site, whose spectral fraction decreases sharply from 16% to 8% as $x$ increases, together with a significant drop in $B_{hyp}$ (~10%). This indicates that the local magnetic field sensed by $Fe^{3+}$ at 4$f$ is strongly perturbed by $Al^{3+}$ substitution at neighboring sites, consistent with the breakdown of Fe–O–Fe exchange bridges linking 4$f$ with 12$k$ and 2$a$ sublattices (**Figure 2d**). The 2$a$ (octahedral) site shows a nearly constant relative area (~10–12%) but a substantial reduction in $B_{hyp}$ (~10%), which, despite partial spectral overlap with the octahedral 4$f$ component, confirms that $Al^+$ preferentially occupies the 2$a$ position. The hyperfine field at 4$f$ (octahedral) also decreases slightly (from 51.7 to 50.0 T), in line with the progressive weakening of the superexchange network.

Although neutron diffraction identifies the 4$e$ site as the one exhibiting the largest decrease in ordered magnetic moment, Mössbauer spectrometry highlights the tetrahedral 4$f$ site as the most affected in terms of local $B_{hyp}$. We believe that these two findings are not contradictory; rather, they are complementary, jointly indicating that $Al^+$ substitution weakens the superexchange backbone, leading simultaneously to the loss of magnetic order at 4$e$ and to a strong reduction of the transferred $B_{hyp}$ at 4$f$. The tetrahedral 4$f$ site is tightly coupled to the spin-up 12$k$ and 2$a$ octahedral sublattices through Fe–O–Fe superexchange bridges that connect the S and R blocks of SFO. As $Al^{3+}$ progressively substitutes $Fe^{3+}$ in these exchange-active sites, the corresponding Al–O–Fe linkages transmit weaker magnetic fields, thereby markedly reducing the transferred $B_{hyp}$ at the tetrahedral 4$f$. In contrast, the 4$e$ trigonal bipyramidal site (located within the Sr-containing R block) is linked through only a few Fe–O–Fe bridges, mainly to 12$k$ across the R–S interface and to the octahedral 4$f$ within the same block. When these neighboring sites lose moment upon Al substitution, the limited exchange connectivity of 4$e$ makes it particularly sensitive, resulting in the strongest reduction of its ordered moment observed by neutron diffraction. In summary, these results show that $Al^{3+}$ substitution weakens the magnetic superexchange network through two coupled effects: a local reduction in transferred hyperfine field at the strongly exchange-connected 4$f$ site, and a progressive destabilization of long-range order at the more weakly coupled 4$e$ site.

*Coercivity Enhancement*

Next, we recall the remarkable increase in coercive field exhibited by the Al-substituted samples. Theoretical studies by Dixit et al.[45] predict that Al substitution enhances the anisotropy field and potentially increases coercivity, depending on the site occupancy. Experimentally, SFAO_x has demonstrated exceptional values of $H_C$ and anisotropy field ($H_A$)[50,51], possibly exceeding those of NdFeB magnets[52], with coercivities up to 1.5 T and $H_A$ values as high as 2.62 MA·m$^{-1}$ in SFAO_2 single crystals[53]. While several studies have attributed such effects to changes in the local superexchange network (particularly around the bipyramidal 4$e$ site), our data suggest a different origin. In our case, the platelet-like morphology is preserved across the series, and the small variation in aspect ratio is insufficient to explain the large differences in coercivity [16,54]. It was shown that the key factor appears to be the evolution of the magnetic domain regime[18]. The coercive field increases systematically with $Al^{3+}$ substitution (Figure S2), reaching values as high as 945 kA/m ($\mu_0 H_C \approx 1.2$ T) for SFAO_2.4, more than twice that of unsubstituted SFO and among the largest reported in the literature[7,55]. This increase correlates directly with the behaviour of the Hopkinson peak observed in the temperature-dependent susceptibility (**Figure 3a**). The much weaker and flatter Hopkinson peak in SFO reflects the dominance of multidomain behaviour, where domain wall motion dominates the susceptibility response rather than coherent rotation. This interpretation is fully consistent with the estimated critical diameters for single-domain behavior[18,56].

*Temperature dependent physical properties*

The evolution of the Curie temperature with Al substitution was assessed from the temperature dependence of the bulk mass susceptibility ($\chi_{mass}$), shown in **Figure 3a**. For all compositions, a well-defined Hopkinson peak marks the ferrimagnetic-to-paramagnetic transition, and its maximum was taken as the experimental Curie temperature, $T_C$. This approach is commonly employed, as the Hopkinson peak provides a sharp and reproducible indicator of the loss of long-range ferrimagnetic order[57]). The extracted values of $T_C$ are reported in **Figure 3b**. The unsubstituted SFO sample exhibits a $T_C$ close to the expected value for $SrFe_{12}O_{19}$, while progressive substitution by $Al^{3+}$ leads to a systematic reduction of $T_C$. This trend directly reflects the weakening of Fe–O–Fe superexchange interactions when nonmagnetic Al replaces $Fe^{3+}$ in the spin-up octahedral $2a$ and $12k$ sites, as established in the Rietveld refinements. For instance, a decrease of almost 100 K is observed between SFO and SFAO_2.4, consistent with the substantial loss of spin-up exchange interactions at higher substitution levels. Quantitatively, this corresponds to an average slope of ~40 K per Al per formula unit, which closely matches the calculated reduction of the net magnetic moment obtained from the refined site occupancies of the $2a$ and $12k$ positions. Moreover, while the $4e$ (bipyramidal) site is only moderately substituted, its refined magnetic moment also decreases significantly with Al content. This was explained in the previous section by the weakening of superexchange links with $12k$ and $4f$ sites, which indirectly destabilizes the $4e$ contribution and further reduces the overall exchange stiffness. Thus, the reduction of $T_C$ reflects not only the direct dilution of the spin-up sublattice but also the indirect destabilization of the bipyramidal $4e$ site. Importantly, the calculated slope is in very good agreement with previous reports on $(Ba/Sr)(Fe_{12-x}Al_xO_{19})$, where similar values of 35–45 K per Al per formula unit were observed[11,58–61], confirming the substitution–$T_C$ correlation across different M-type hexaferrites. To further probe the evolution of the ferrimagnetic structure across the transition regime, we performed NPD measurements on selected samples at different temperatures. **Figure 4a** shows a detail of the 100 reflections of the SFO sample collected at 15, 298, 500, and 800 K. As the temperature increases, the magnetic contribution to the reflection systematically decreases, vanishing above the $T_C$, in agreement with the susceptibility measurements discussed above. For the unsubstituted SFO, the low-temperature measurement at 15 K confirms a slightly larger ordered moment compared to 298 K. However, the difference is marginal, indicating that thermal agitation only induces a small reduction of the ordered spin alignment at room temperature. For this reason, structural refinements were focused on the data collected at 298 K for direct comparison with the Al-substituted samples.

**Figure 4b** reports the average magnetic moment per $Fe^{3+}$ ion as a function of temperature, extracted from Rietveld refinements using the NPD patterns. The data clearly shows a decay with increasing temperature, approaching zero at the Curie point. The dashed vertical lines in **Figure 4b** correspond to the $T_C$ values independently determined from bulk susceptibility (**Figures 3a,b**), highlighting the excellent consistency between the two techniques. The dotted curves represent exponential fits to the temperature dependence of the average moment, plotted as a guide to the eye. Notably, the moment values obtained after cooling back to 298 K from

800 K (open markers) remain within the experimental error of the initial 298 K data, ruling out irreversible thermal effects under the investigated conditions. This combined analysis of NPD and susceptibility measurements provides a coherent picture of the progressive loss of long-range ferrimagnetic order with increasing temperature.

Raman spectroscopy was employed to gain further insight into the lattice dynamics and their correlation with the magnetic transition at high T. **Figure 5a** shows the Raman spectrum of SFO collected at 300 K, in the frequency range of 100 – 900 cm$^{-1}$, where the experimental data were fitted using pseudo-Voigt functions to resolve the individual contributions (total twelve modes). Figure S4 shows the full set of spectra. SFO belong to $P6_3/mmc$ space group at room temperature where group theory predicts 42 Raman active phonons, which can be characterized according to the irreducible presentation as $11A_{1g} + 14E_{1g} + 17E_{2g}$. The assignment of the observed modes is consistent with previous reports for hexagonal M-type ferrites, with active vibrations arising from FeO$_6$ octahedra, FeO$_4$ tetrahedra, and FeO$_5$ bipyramids[62]. No additional peaks have been observed which further confirm singular phase of these composites. These Raman modes are assigned to vibration modes arising from different chemical environments of Fe$^{3+}$. Specifically, the peak observed at 410 cm$^{-1}$ is assigned to the octahedral sites ($12k$-dominated, and above $12k/2a$ mixed), the peak at 610 cm$^{-1}$ corresponds to the $4f$ octahedral site and the peak at 684 cm$^{-1}$ is attributed to the $4e$ bipyramidal site sublattice vibration modes, respectively, characteristic of strontium hexaferrite[63,64]. Lower-frequency modes (<200 cm$^{-1}$) correspond mainly to the whole spinel block[62]. Several modes are observed in the range 300 - 600 cm$^{-1}$ coming from three different octahedral groups due to metal-oxygen (M - O) stretching vibrations[62]. Similar results were also observed in various spinel structures containing iron cations in the octahedral sites[65]. Comparison of the Raman spectra with the spinel structure is reasonable because the complex crystal structure of hexaferrite consists of periodic stacks of spinel (Fe$_6$O$_8$$^{2+}$, S-block) and hexagonal sub-structures (SrFe$_6$O$_{11}$$^{2-}$, R-block)[66]. The mode for $4f$ tetrahedral site is overlapped (~720 cm$^{-1}$) with the stronger A$_{1g}$-bipyramid at 684 cm$^{-1}$. Peak positions of all phonon vibrations distinguished from fitted curves are summarized in Table S6 (a comparison of variational modes with other hexaferrites are also presented).

The computationally calculated phonon modes of SFO reproduce the expected irreducible representation, and the computed Raman spectrum captures the main experimental features (see Figures 5c and S4). The most intense calculated mode appears near 678 cm$^{-1}$ and is attributed to the dominant band observed experimentally. Raman-active phonons extend from ~180 cm$^{-1}$ to ~750 cm$^{-1}$, in good agreement with the measured spectra.

The comparison between experimental SFO and Al-substituted SFAO_2 (**Figure 5a,b**) reveals a blue shift of the modes upon increasing Al content, consistent with the reduced lattice parameters obtained from Rietveld refinements. Figure S6 show the full set of spectra, showing also a few modes for example, at 684, 610, 404.5 and 334.6 cm$^{-1}$ which show relatively larger shift in peak positions with Al substitution: these modes shift more than 5% as the Al doping increases up to 2.4. To rationalize these trends, first-principles calculations were performed for two possible Al substitutional sites ($2a$ and $12f$). The 2a site is energetically favored by 0.07 eV, and the theoretical spectrum discussed in **Figure 5c** therefore corresponds to Al occupying the $2a$ position. The calculations show that Al incorporation induces a clear splitting of the Raman features, accompanied by a positive shift (e.g. A$_{1g}$ at 686 cm$^{-1}$ is shifted of ~7 cm$^{-1}$, see also Figure S7). Mode visualizations (Figure S8,9) confirm that partial substitution perturbs the local vibrational environment, giving rise to the observed splitting. Additional Raman modes are also modified by Al incorporation: the ~604 cm$^{-1}$ mode shifts upward and becomes split, the ~531 cm$^{-1}$ mode develops upward and downward components, and the ~289 cm$^{-1}$ mode is similarly perturbed, all in agreement with experimental trends.

The evolution of the Raman spectra with temperature is shown in **Figure 5d.** Upon heating, all phonon modes gradually broaden and soften (red shift), reflecting the combined effects of lattice expansion and anharmonic phonon–phonon interactions (as captured by the Balkanski model)[67]. However, near the ferrimagnetic-to-paramagnetic transition, additional anomalies become evident, suggesting a coupling between the spin and lattice degrees of freedom[68].

To quantify these effects, **Figure 6** compares the temperature dependence of peak position, and full width at half maximum (FWHM) for selected modes ($A_{1g}$-bypiramid and $A_{1g}$-octahedra) of SFO, SFAO_1.4, and SFAO_2. The frequency shifts of representative phonons were analysed using the Balkanski model, which describes anharmonic decay processes of optical phonons. While the data at low and intermediate temperatures can be satisfactorily fitted within this framework, deviations appear close to the T$_C$. In this region, the observed phonon softening is stronger than predicted by purely anharmonic effects, highlighting the presence of spin–phonon coupling[69,70]. It is well known that the FWHM is related to the phonon lifetime[71]: a slight change in the lattice due to the coupling changes phonons lifetime and, consequently, the FWHM. In a similar manner such effect is also observed near to magnetic transitions, which further confirms strong spin-phonon coupling at high temperatures. From the observed trend, the bipyramidal Fe–O modes exhibit the strongest anomalies, as their effective bond stiffness is highly sensitive to the surrounding spin configuration and is destabilized by the weakening of the 4$e$–12$k$ superexchange network. This behavior is fully consistent with the NPD and Mössbauer analyses, which reveal a pronounced reduction of the magnetic moment at the *4e* site despite the absence of direct Al substitution, confirming that the disruption of the interactions is the key origin of the phonon anomalies. Importantly, the temperatures at which these anomalies occur (i.e., the clear slope change around 750, 665 and 630 K for SFO, SFAO_1.4 and SFAO_2) correlate well with the T$_C$ independently determined from magnetic susceptibility measurements (**Figures 3a,b**), confirming that the phonon dynamics are directly sensitive to the magnetic ordering transition.

The magnitude of these effects is found to depend on the Al content. In particular, samples with higher Al substitution exhibit shows a stronger phonon softening and linewidth broadening near T$_C$, which can be rationalized in terms of the progressive weakening of Fe–O–Fe superexchange interactions (as they contribute to the effective spring constant of Fe–O bonds) and the enhanced sensitivity of the lattice to magnetic fluctuations when non-magnetic Al replaces Fe$^{3+}$.

Given that Raman spectroscopy already revealed how Al substitution affects the lattice dynamics through phonon softening and broadening near T$_C$, it is natural to extend the analysis to the dielectric properties, which provide complementary insight into how charge and dipolar dynamics couple to the evolving magnetic and structural environment. The dielectric properties of SrFe$_{12-x}$Al$_x$O$_{19}$ $x$ = 0, 1, 1.4, 2 have been collected, and the relative dielectric constant and loss are depicted in **Figures 7a,b**. While some systematics can be observed, *e.g.* the overall reduction of ε$_r$' and tanδ near room temperature with increasing Al content, the overall dielectric response appears to be a collection of overlapping relaxations. To get some more insight on the dielectric properties, we have evaluated the ac-conductivity, σ$_{ac}$ = ωε$_0$ε$_r$' tanδ, electric modulus

M*=1/ε* and impedance Z* = 1/iωC₀ε* (more details in Ref.[72]). As seen in **Figure 7c**, the ac-conductivity is found to be significantly decreasing as the Al content increases from 0 to 1.4. It increases again for x = 2, in agreement with the loss data presented in **Figure 7b**. The frequency dependence of the dielectric constant and impedance was analysed using an equivalent circuit model, as illustrated in **Figure 7d,e**. The variation of $\varepsilon_r'$ and Z" exhibits a well-defined plateau extending from approximately 50 kHz toward lower frequencies, indicative of Maxwell–Wagner interfacial polarization arising from charge accumulation at the grain boundaries. This observation is further substantiated by the Nyquist plot presented in **Figure 7d**, which reveals two distinct semicircular arcs associated with the grain interior and grain boundary responses, respectively. Inside the grains, Al is found to reduce the number of polaronic $Fe^{2+}/Fe^{3+}$ hopping sites, reducing the carrier density, as also observed in other systems[73]. This, in turn, decreases the overall conductivity of the Al-substituted samples and their dielectric losses.

Taken together, these results suggest that while neutron diffraction, Mössbauer spectrometry, Raman spectroscopy, and dielectric measurements consistently highlight the weakening of Fe–O–Fe exchange and the loss of magnetic moment with Al substitution, the coercivity data demonstrate that at the mesoscale the dominant effect is the stabilization of single-domain reversal (**Figure 8a,b**). This explains how exceptionally hard ferrimagnetic behavior emerges despite the intrinsic weakening of the exchange network.

**Conclusions**

This unified picture highlights the central role of superexchange weakening in governing both the magnetic and functional responses of Al-substituted $SrFe_{12}O_{19}$. The combined NPD and XRPD refinements show that $Al^{3+}$ preferentially replaces $Fe^{3+}$ in spin-up octahedral sites (2a, 12k), leading to a systematic reduction of the net magnetic moment and Curie temperature. Complementary $^{57}Fe$ Mössbauer spectrometry confirms this preferential substitution, revealing a marked decrease of the hyperfine field at 2a and tetrahedral 4f sites, consistent with the weakening of the transferred exchange field within the spin-up sublattice. Raman spectroscopy further reveals pronounced phonon softening and linewidth broadening near $T_C$, with the strongest anomalies linked to bipyramidal Fe–O vibrations, consistent with the disruption of the 4e–12k/4e–4f superexchange network. The dielectric measurements suggest that the Al incorporation modifies the polaronic conduction across the Al substituted Fe-O-Fe bonds, leading to the enhancement of the capacitive behaviour. At the mesoscale, susceptibility and coercivity data demonstrates that Al substitution stabilizes a single-domain reversal regime, leading to a substantial increase of coercive field up to ~1.2 T. Altogether, these findings reconcile the apparent paradox of reduced exchange interactions and enhanced hard magnetic performance, highlighting the interplay between atomic-scale substitution effects and domain-scale reversal mechanisms in determining the functional properties of M-type hexaferrites.

**Acknowledgements**

This work was supported by the Swedish Energy Agency (project numbers 46561-1 and 50667-1), the Swedish Research Council (VR 2023-05244) and the OlleEngkvist Stiftelse (Grant No. 224-0046). It has also received funding from the European Union's Horizon 2020 research and innovation program under grant agreement n°823717 – ESTEEM3 (Enabling Science and Technology through European Electron Microscopy). The authors acknowledge financial support from the MEYS CR (Project OP JAK FerrMion, No. CZ.02.01.01/00/22_008/0004591). Neutron diffraction measurements were carried out at the CANAM infrastructure of the NPI CAS Rez, using the CICRR infrastructure supported by MEYS project LM2023041, both of which are acknowledged. This work was partially developed in the framework of the PRIN 2022-PNRR project HyperMag (prot. P2022RRRT4, CUP Unige: D53D23017200001; CUP Ca' Foscari: H53D23007990001) and supported by European Union – NextGenerationEU, and by "Network 4 Energy Sustainable Transition-NEST" project (code PE0000021), adopted by the "Ministero dell'Università e della Ricerca (MUR)," according to attachment E of Decree No. 1561/2022. Investigation conducted by S.A.I. was supported by the State budget for scientific research at Lomonosov Moscow State University (project № 121031300090-2).

**ORCID**

Pierfrancesco Maltoni 0000-0001-9834-3164
Ravi K. Dokala 0000-0002-3466-8717
Prativa Pramanik
Rafael Araujo 0000-0002-5261-2047
Tomas Edvinsson 0000-0003-2759-7356
Sergey A. Ivanov 0000-0001-6073-0708
Bjarne Almqvist 0000-0002-9385-7614
Gaspare Varvaro 0000-0001-7313-7268
Aldo Capobianchi 0000-0003-3956-2062
Nader Yaacoub 0000-0002-5583-1165
Charles Hervoches 0000-0002-9550-3289
Alberto Martinelli 0000-0001-8391-3486
Robert C. Pullar 0000-0001-6844-4482
Davide Peddis 0000-0003-0810-8860
Roland Mathieu 0000-0002-5261-2047

**References**

(1) Jones, N. Materials Science: The Pull of Stronger Magnets. *Nature* 2011, *472* (7341), 22–23. https://doi.org/10.1038/472022a.
(2) Gutfleisch, O.; Willard, M. A.; Brück, E.; Chen, C. H.; Sankar, S. G.; Liu, J. P. Magnetic Materials and Devices for the 21st Century: Stronger, Lighter, and More Energy Efficient. *Advanced Materials* 2011, *23* (7), 821–842. https://doi.org/10.1002/adma.201002180.
(3) Müller, K.-H.; Sawatzki, S.; Gauß, R.; Gutfleisch, O. Permanent Magnet Materials and Applications. In *Handbook of Magnetism and Magnetic Materials*; Springer International Publishing: Cham, 2021; pp 1369–1433. https://doi.org/10.1007/978-3-030-63210-6_29.
(4) Golroudbary, S. R.; Makarava, I.; Kraslawski, A.; Repo, E. Global Environmental Cost of Using Rare Earth Elements in Green Energy Technologies. *Science of The Total Environment* 2022, *832*, 155022. https://doi.org/10.1016/j.scitotenv.2022.155022.
(5) Schönfeldt, M.; Opelt, K.; Hasan, M.; Gröninger, M.; Jahnke, D.; Gassmann, J.; Gutfleisch, O. Functional Recycling and Reuse of Nd–Fe–B Permanent Magnets from Various Waste Streams for a More Sustainable and Resilient Electromobility. *Adv Eng Mater* 2025, *27* (7). https://doi.org/10.1002/adem.202402815.

(6) Saura-Múzquiz, M.; Granados-Miralles, C.; Andersen, H. L.; Stingaciu, M.; Avdeev, M.; Christensen, M. Nanoengineered High-Performance Hexaferrite Magnets by Morphology-Induced Alignment of Tailored Nanoplatelets. *ACS Appl Nano Mater* 2018, *1* (12), 6938–6949. https://doi.org/10.1021/acsanm.8b01748.

(7) de Julian Fernandez, C.; Sangregorio, C.; de la Figuera, J.; Belec, B.; Makovec, D.; Quesada, A. Topical Review: Progress and Prospects of Hard Hexaferrites for Permanent Magnet Applications. *J Phys D Appl Phys* 2020. https://doi.org/10.1088/1361-6463/abd272.

(8) Bollero, A.; Palmero, E. M. Recent Advances in Hard -ferrite Magnets. In *Modern Permanent Magnets*; Elsevier, 2022; pp 65–112. https://doi.org/10.1016/B978-0-323-88658-1.00013-3.

(9) Obradors, X.; Solans, X.; Collomb, A.; Samaras, D.; Rodriguez, J.; Pernet, M.; Font-Altaba, M. Crystal Structure of Strontium Hexaferrite $SrFe_{12}O_{19}$. *J Solid State Chem* 1988, *72* (2), 218–224. https://doi.org/10.1016/0022-4596(88)90025-4.

(10) Kojima, H. Chapter 5 Fundamental Properties of Hexagonal Ferrites with Magnetoplumbite Structure; 1982; pp 305–391. https://doi.org/10.1016/S1574-9304(05)80091-4.

(11) Pullar, R. C. Hexagonal Ferrites: A Review of the Synthesis, Properties and Applications of Hexaferrite Ceramics. *Prog Mater Sci* 2012, *57* (7), 1191–1334. https://doi.org/10.1016/j.pmatsci.2012.04.001.

(12) Mahmood, S.; Abu-Aljarayesh, I. *Hexaferrite Permanent Magnetic Materials*; Materials Research Forum LLC, 2016; Vol. 4. https://doi.org/10.21741/9781945291074.

(13) Lisjak, D.; Drofenik, M. Chemical Substitution—An Alternative Strategy for Controlling the Particle Size of Barium Ferrite. *Cryst Growth Des* 2012, *12* (11), 5174–5179. https://doi.org/10.1021/cg301227r.

(14) Singh, J.; Singh, C.; Kaur, D.; Zaki, H.; Abdel-Latif, I. A.; Narang, S. B.; Jotania, R.; Mishra, S. R.; Joshi, R.; Dhruv, P.; Ghimire, M.; Shirsath', S. E.; Meena, S. S. Elucidation of Phase Evolution, Microstructural, Mössbauer and Magnetic Properties of $Co^{2+}Al^{3+}$ Doped M-Type Ba Sr Hexaferrites Synthesized by a Ceramic Method. *J Alloys Compd* 2017, *695*, 1112–1121. https://doi.org/10.1016/j.jallcom.2016.10.237.

(15) Dixit, V.; Nandadasa, C. N.; Kim, S.-G.; Kim, S.; Park, J.; Hong, Y.-K.; Liyanage, L. S. I.; Moitra, A. Site Occupancy and Magnetic Properties of Al-Substituted M-Type Strontium Hexaferrite. *J Appl Phys* 2015, *117* (24). https://doi.org/10.1063/1.4922867.

(16) Stingaciu, M.; Mishra, D.; de Julián Fernández, C.; Cabassi, R.; Eikeland, A. Z.; Christensen, M.; Deledda, S. High Magnetic Coercive Field in Ca-Al-Cr Substituted Strontium Hexaferrite. *J Alloys Compd* 2021, *883*, 160768. https://doi.org/10.1016/j.jallcom.2021.160768.

(17) Xie, B.; Zhou, X.; Chen, W.; Fan, L.; Zhang, L.; Li, R.; Zheng, H.; Wu, Q.; Wu, Y.; Lin, Y.; Zheng, P.; Zheng, L.; Zhang, Y. High Remanence Ratio of Aluminum Substituted Hexagonal Barium Ferrite Films for Self-Biased Microwave Devices. *J Alloys Compd* 2023, *938*, 168710. https://doi.org/10.1016/j.jallcom.2023.168710.

(18) Maltoni, P.; Barucca, G.; Rutkowski, B.; Ivanov, S. A.; Yaacoub, N.; Mikheenkova, A.; Ek, G.; Eriksson, M.; Almqvist, B.; Vasilakaki, M.; Varvaro, G.; Sarkar, T.; De Toro, J. A.; Trohidou, K.; Peddis, D.; Mathieu, R. Engineering Hard Ferrite Composites by Combining Nanostructuring and $Al^{3+}$ Substitution: From Nano to Dense Bulk Magnets. *Acta Mater* 2025, *282*, 120491. https://doi.org/10.1016/j.actamat.2024.120491.

(19) Saura-Múzquiz, M.; Eikeland, A. Z.; Stingaciu, M.; Andersen, H. L.; Avdeev, M.; Christensen, M. Synthesis, Crystal Structure, Site Occupancy, and Magnetic Properties of Aluminum-Substituted M-Type Sr Hexaferrite $SrFe_{12-x}Al_xO_{19}$ Nanoparticles. *Chemistry of Materials* 2025, *37* (3), 884–896. https://doi.org/10.1021/acs.chemmater.4c02205.

(20) Rhein, F.; Helbig, T.; Neu, V.; Krispin, M.; Gutfleisch, O. In-Situ Magnetic Force Microscopy Analysis of Magnetization and Demagnetization Behavior in $Al^{3+}$ Substituted Sr-Hexaferrite. *Acta Mater* 2018, *146*, 85–96. https://doi.org/10.1016/j.actamat.2017.12.010.

(21) Marek, E.; Hu, W.; Gaultois, M.; Grey, C. P.; Scott, S. A. The Use of Strontium Ferrite in Chemical Looping Systems. *Appl Energy* 2018, *223*, 369–382. https://doi.org/10.1016/j.apenergy.2018.04.090.
(22) Fasil, M.; Mijatovic, N.; Jensen, B. B.; Holboll, J. Performance Variation of Ferrite Magnet PMBLDC Motor With Temperature. *IEEE Trans Magn* 2015, *51* (12), 1–6. https://doi.org/10.1109/TMAG.2015.2456854.
(23) Shanshal, A.; Hoang, K.; Atallah, K. High-Performance Ferrite Permanent Magnet Brushless Machines. *IEEE Trans Magn* 2019, *55* (7), 1–4. https://doi.org/10.1109/TMAG.2019.2900561.
(24) Sugimoto, M. The Past, Present, and Future of Ferrites. *Journal of the American Ceramic Society* 1999, *82* (2), 269–280. https://doi.org/10.1111/j.1551-2916.1999.tb20058.x.
(25) Wartewig, P.; Krause, M. K.; Esquinazi, P.; Rösler, S.; Sonntag, R. Magnetic Properties of Zn- and Ti-Substituted Barium Hexaferrite. *J Magn Magn Mater* 1999, *192* (1), 83–99. https://doi.org/10.1016/S0304-8853(98)00382-5.
(26) Ben Ghzaiel, T.; Dhaoui, W.; Pasko, A.; Mazaleyrat, F. Effect of Non-Magnetic and Magnetic Trivalent Ion Substitutions on BaM-Ferrite Properties Synthesized by Hydrothermal Method. *J Alloys Compd* 2016, *671*, 245–253. https://doi.org/10.1016/j.jallcom.2016.02.071.
(27) Tenaud, P.; Morel, A.; Kools, F.; Le Breton, J. M.; Lechevallier, L. Recent Improvement of Hard Ferrite Permanent Magnets Based on La–Co Substitution. *J Alloys Compd* 2004, *370* (1–2), 331–334. https://doi.org/10.1016/j.jallcom.2003.09.106.
(28) Ueda, H.; Tanioku, Y.; Michioka, C.; Yoshimura, K. Magnetocrystalline Anisotropy of La- and Co-Substituted M-Type Strontium Ferrites: Role of $Co^{2+}$ and $Fe^{2+}$. *Phys Rev B* 2017, *95* (22), 224421. https://doi.org/10.1103/PhysRevB.95.224421.
(29) Kumar, S.; Supriya, S.; Kar, M. Correlation between Temperature Dependent Dielectric and DC Resistivity of Cr Substituted Barium Hexaferrite. *Mater Res Express* 2017, *4* (12), 126302. https://doi.org/10.1088/2053-1591/aa9a51.
(30) Lechevallier, L.; Le Breton, J. M.; Morel, A.; Tenaud, P. On the Solubility of Rare Earths in M-Type $SrFe_{12}O_{19}$ Hexaferrite Compounds. *Journal of Physics: Condensed Matter* 2008, *20* (17), 175203. https://doi.org/10.1088/0953-8984/20/17/175203.
(31) Maltoni, P.; Sarkar, T.; Barucca, G.; Varvaro, G.; Peddis, D.; Mathieu, R. Exploring the Magnetic Properties and Magnetic Coupling in $SrFe_{12}O_{19}/Co_{1-x}Zn_xFe_2O_4$ Nanocomposites. *J Magn Magn Mater* 2021, *535*, 168095. https://doi.org/10.1016/j.jmmm.2021.168095.
(32) Dirba, I.; Mohammadi, M.; Rhein, F.; Gong, Q.; Yi, M.; Xu, B.-X.; Krispin, M.; Gutfleisch, O. Synthesis and Magnetic Properties of Bulk $A''-Fe_{16}N_2/SrAl_2Fe_{10}O_{19}$ Composite Magnets. *J Magn Magn Mater* 2021, *518*, 167414. https://doi.org/10.1016/j.jmmm.2020.167414.
(33) Rietveld, H. M. A Profile Refinement Method for Nuclear and Magnetic Structures. *J Appl Crystallogr* 1969, *2* (2), 65–71. https://doi.org/10.1107/S0021889869006558.
(34) Rodríguez-Carvajal, J. Recent Advances in Magnetic Structure Determination by Neutron Powder Diffraction. *Physica B Condens Matter* 1993, *192* (1–2), 55–69. https://doi.org/10.1016/0921-4526(93)90108-I.
(35) Pramanik, P.; Eder, F.; Weil, M.; Ivanov, S. A.; Maltoni, P.; Miletich, R.; Edvinsson, T.; Mathieu, R. Vibrational Properties of Monoclinic $CoTeO_4$. *Phys Rev B* 2024, *110* (5), 054104. https://doi.org/10.1103/PhysRevB.110.054104.
(36) Kresse, G.; Furthmüller, J. Efficient Iterative Schemes for *Ab Initio* Total-Energy Calculations Using a Plane-Wave Basis Set. *Phys Rev B* 1996, *54* (16), 11169–11186. https://doi.org/10.1103/PhysRevB.54.11169.
(37) Kresse, G.; Joubert, D. From Ultrasoft Pseudopotentials to the Projector Augmented-Wave Method. *Phys Rev B* 1999, *59* (3), 1758–1775. https://doi.org/10.1103/PhysRevB.59.1758.
(38) Perdew, J. P.; Burke, K.; Ernzerhof, M. Generalized Gradient Approximation Made Simple. *Phys Rev Lett* 1996, *77* (18), 3865–3868. https://doi.org/10.1103/PhysRevLett.77.3865.

(39) Dudarev, S. L.; Botton, G. A.; Savrasov, S. Y.; Humphreys, C. J.; Sutton, A. P. Electron-Energy-Loss Spectra and the Structural Stability of Nickel Oxide: An LSDA+U Study. *Phys Rev B* 1998, *57* (3), 1505–1509. https://doi.org/10.1103/PhysRevB.57.1505.

(40) Hou, Y. H.; Chen, X.; Guo, X. L.; Li, W.; Huang, Y. L.; Tao, X. M. Effects of Intrinsic Defects and Doping on $SrFe_{12}O_{19}$: A First-Principles Exploration of the Structural, Electronic and Magnetic Properties. *J Magn Magn Mater* 2021, *538*, 168257. https://doi.org/10.1016/j.jmmm.2021.168257.

(41) Rhein, F.; Karmazin, R.; Krispin, M.; Reimann, T.; Gutfleisch, O. Enhancement of Coercivity and Saturation Magnetization of $Al^{3+}$ Substituted M-Type Sr-Hexaferrites. *J Alloys Compd* 2017, *690*, 979–985. https://doi.org/10.1016/j.jallcom.2016.08.085.

(42) Kazin, P. E.; Trusov, L. A.; Zaitsev, D. D.; Tretyakov, Yu. D.; Jansen, M. Formation of Submicron-Sized $SrFe_{12-x}Al_xO_{19}$ with Very High Coercivity. *J Magn Magn Mater* 2008, *320* (6), 1068–1072. https://doi.org/10.1016/j.jmmm.2007.10.020.

(43) Gorbachev, E. A.; Lebedev, V. A.; Kozlyakova, E. S.; Alyabyeva, L. N.; Ahmed, A.; Cervellino, A.; Trusov, L. A. Tuning the Microstructure, Magnetostatic and Magnetodynamic Properties of Highly Al-Substituted M-Type Sr/Ca Hexaferrites Prepared by Citrate-Nitrate Auto-Combustion Method. *Ceram Int* 2023. https://doi.org/10.1016/j.ceramint.2023.05.177.

(44) Trusov, L. A.; Gorbachev, E. A.; Lebedev, V. A.; Sleptsova, A. E.; Roslyakov, I. V.; Kozlyakova, E. S.; Vasiliev, A. V.; Dinnebier, R. E.; Jansen, M.; Kazin, P. E. Ca–Al Double-Substituted Strontium Hexaferrites with Giant Coercivity. *Chemical Communications* 2018, *54* (5), 479–482. https://doi.org/10.1039/C7CC08675J.

(45) Dixit, V.; Nandadasa, C. N.; Kim, S.-G.; Kim, S.; Park, J.; Hong, Y.-K.; Liyanage, L. S. I.; Moitra, A. Site Occupancy and Magnetic Properties of Al-Substituted M-Type Strontium Hexaferrite. *J Appl Phys* 2015, *117* (24). https://doi.org/10.1063/1.4922867.

(46) Fang, C. M.; Kools, F.; Metselaar, R.; With, G. de; Groot, R. A. de. Magnetic and Electronic Properties of Strontium Hexaferrite $SrFe_{12}O_{19}$ from First-Principles Calculations. *Journal of Physics: Condensed Matter* 2003, *15* (36), 6229–6237. https://doi.org/10.1088/0953-8984/15/36/311.

(47) Awawdeh, M.; Bsoul, I.; Mahmood, S. H. Magnetic Properties and Mössbauer Spectroscopy on Ga, Al, and Cr Substituted Hexaferrites. *J Alloys Compd* 2014, *585*, 465–473. https://doi.org/10.1016/j.jallcom.2013.09.174.

(48) Trukhanov, A. V.; Kostishyn, V. G.; Panina, L. V.; Jabarov, S. H.; Korovushkin, V. V.; Trukhanov, S. V.; Trukhanova, E. L. Magnetic Properties and Mössbauer Study of Gallium Doped M-Type Barium Hexaferrites. *Ceram Int* 2017, *43* (15), 12822–12827. https://doi.org/10.1016/j.ceramint.2017.06.172.

(49) Trukhanov, S. V.; Trukhanov, A. V.; Turchenko, V. A.; Kostishyn, V. G.; Panina, L. V.; Kazakevich, I. S.; Balagurov, A. M. Structure and Magnetic Properties of $BaFe_{11.9}In_{0.1}O_{19}$ Hexaferrite in a Wide Temperature Range. *J Alloys Compd* 2016, *689*, 383–393. https://doi.org/10.1016/j.jallcom.2016.07.309.

(50) Gorbachev, E. A.; Kozlyakova, E. S.; Alyabyeva, L. N.; Ahmed, A.; Trusov, L. A. Hard Ferrite Magnetic Insulators Revealing Giant Coercivity and Sub-Terahertz Natural Ferromagnetic Resonance at 5–300 K. *Mater Horiz* 2023, *10* (5), 1842–1847. https://doi.org/10.1039/D3MH00089C.

(51) Gorbachev, E. A.; Trusov, L. A.; Sleptsova, A. E.; Kozlyakova, E. S.; Alyabyeva, L. N.; Yegiyan, S. R.; Prokhorov, A. S.; Lebedev, V. A.; Roslyakov, I. V.; Vasiliev, A. V.; Kazin, P. E. Hexaferrite Materials Displaying Ultra-High Coercivity and Sub-Terahertz Ferromagnetic Resonance Frequencies. *Materials Today* 2020, *32*, 13–18. https://doi.org/10.1016/j.mattod.2019.05.020.

(52) Lewis, L. H.; Jiménez-Villacorta, F. Perspectives on Permanent Magnetic Materials for Energy Conversion and Power Generation. *Metallurgical and Materials Transactions A* 2013, *44* (S1), 2–20. https://doi.org/10.1007/s11661-012-1278-2.

(53) Ustinov, A. B.; Tatarenko, A. S.; Srinivasan, G.; Balbashov, A. M. Al Substituted Ba-Hexaferrite Single-Crystal Films for Millimeter-Wave Devices. *J Appl Phys* 2009, *105* (2). https://doi.org/10.1063/1.3067759.

(54) Abo, G. S.; Hong, Y.-K.; Jalli, J.; Lee, J.-J.; Park, J.-H.; Bae, S.; Kim, S.-G.; Choi, B.-C.; Tanaka, T. Shape Dependent Coercivity Simulation of a Spherical Barium Ferrite (S-BaFe) Particle with Uniaxial Anisotropy. *Journal of Magnetics* 2012, *17* (1), 1–5. https://doi.org/10.4283/JMAG.2012.17.1.001.

(55) Trusov, L. A.; Gorbachev, E. A.; Lebedev, V. A.; Sleptsova, A. E.; Roslyakov, I. V.; Kozlyakova, E. S.; Vasiliev, A. V.; Dinnebier, R. E.; Jansen, M.; Kazin, P. E. Ca–Al Double-Substituted Strontium Hexaferrites with Giant Coercivity. *Chemical Communications* 2018, *54* (5), 479–482. https://doi.org/10.1039/C7CC08675J.

(56) Kittel, C. Physical Theory of Ferromagnetic Domains. *Rev Mod Phys* 1949, *21* (4), 541–583. https://doi.org/10.1103/RevModPhys.21.541.

(57) Pfeiffer, H.; Schüppel, W. Temperature Dependence of the Magnetization in Fine Particle Systems and the Hopkinson Effect. Application to Barium Ferrite Powders. *J Magn Magn Mater* 1994, *130* (1–3), 92–98. https://doi.org/10.1016/0304-8853(94)90661-0.

(58) Trusov, L. A.; Gorbachev, E. A.; Lebedev, V. A.; Sleptsova, A. E.; Roslyakov, I. V.; Kozlyakova, E. S.; Vasiliev, A. V.; Dinnebier, R. E.; Jansen, M.; Kazin, P. E. Ca–Al Double-Substituted Strontium Hexaferrites with Giant Coercivity. *Chemical Communications* 2018, *54* (5), 479–482. https://doi.org/10.1039/C7CC08675J.

(59) Chen, D.; Harward, I.; Baptist, J.; Goldman, S.; Celinski, Z. Curie Temperature and Magnetic Properties of Aluminum Doped Barium Ferrite Particles Prepared by Ball Mill Method. *J Magn Magn Mater* 2015, *395*, 350–353. https://doi.org/10.1016/j.jmmm.2015.07.076.

(60) Kazin, P. E.; Trusov, L. A.; Zaitsev, D. D.; Tretyakov, Yu. D.; Jansen, M. Formation of Submicron-Sized $SrFe_{12-x}Al_xO_{19}$ with Very High Coercivity. *J Magn Magn Mater* 2008, *320* (6), 1068–1072. https://doi.org/10.1016/j.jmmm.2007.10.020.

(61) Luo, H.; Rai, B. K.; Mishra, S. R.; Nguyen, V. V.; Liu, J. P. Physical and Magnetic Properties of Highly Aluminum Doped Strontium Ferrite Nanoparticles Prepared by Auto-Combustion Route. *J Magn Magn Mater* 2012, *324* (17), 2602–2608. https://doi.org/10.1016/j.jmmm.2012.02.106.

(62) Kreisel, J.; Lucazeau, G.; Vincent, H. Raman Spectra and Vibrational Analysis of $BaFe_{12}O_{19}$ Hexagonal Ferrite. *J Solid State Chem* 1998, *137* (1), 127–137. https://doi.org/10.1006/jssc.1997.7737.

(63) Morel, A.; Le Breton, J. M.; Kreisel, J.; Wiesinger, G.; Kools, F.; Tenaud, P. Sublattice Occupation in $Sr_{1-x}La_xFe_{12-x}Co_xO_{19}$ Hexagonal Ferrite Analyzed by Mössbauer Spectrometry and Raman Spectroscopy. *J Magn Magn Mater* 2002, *242–245*, 1405–1407. https://doi.org/10.1016/S0304-8853(01)00962-3.

(64) Soria, G. D.; Marco, J. F.; Mandziak, A.; Sánchez-Cortés, S.; Sánchez-Arenillas, M.; Prieto, J. E.; Dávalos, J.; Foerster, M.; Aballe, L.; López-Sánchez, J.; Guzmán-Mínguez, J. C.; Granados-Miralles, C.; de la Figuera, J.; Quesada, A. Influence of the Growth Conditions on the Magnetism of $SrFe_{12}O_{19}$ Thin Films and the Behavior of Co/ $SrFe_{12}O_{19}$ Bilayers. *J Phys D Appl Phys* 2020, *53* (34), 344002. https://doi.org/10.1088/1361-6463/ab8d70.

(65) Degiorgi, L.; Blatter-Mörke, I.; Wachter, P. Magnetite: Phonon Modes and the Verwey Transition. *Phys Rev B* 1987, *35* (11), 5421–5424. https://doi.org/10.1103/PhysRevB.35.5421.

(66) Maltoni, P.; Sarkar, T.; Varvaro, G.; Barucca, G.; Ivanov, S. A.; Peddis, D.; Mathieu, R. Towards Bi-Magnetic Nanocomposites as Permanent Magnets through the Optimization of the Synthesis and Magnetic Properties of $SrFe_{12}O_{19}$ Nanocrystallites. *J Phys D Appl Phys* 2021, *54* (12), 124004. https://doi.org/10.1088/1361-6463/abd20d.

(67) Balkanski, M.; Wallis, R. F.; Haro, E. Anharmonic Effects in Light Scattering Due to Optical Phonons in Silicon. *Phys Rev B* 1983, *28* (4), 1928–1934. https://doi.org/10.1103/PhysRevB.28.1928.

(68) Buzinaro, M. A. P.; Macêdo, M. A.; Costa, B. F. O.; Ferreira, N. S. Disorder of Fe(2)O$_5$ Bipyramids and Spin-Phonon Coupling in SrFe$_{12}$O$_{19}$ Nanoparticles. *Ceram Int* 2019, *45* (10), 13571–13574. https://doi.org/10.1016/j.ceramint.2019.03.214.

(69) Silva Júnior, F. M.; Paschoal, C. W. A. Spin-Phonon Coupling in BaFe$_{12}$O$_{19}$ M-Type Hexaferrite. *J Appl Phys* 2014, *116* (24). https://doi.org/10.1063/1.4904062.

(70) Shen, S.-P.; Chai, Y.-S.; Cong, J.-Z.; Sun, P.-J.; Lu, J.; Yan, L.-Q.; Wang, S.-G.; Sun, Y. Magnetic-Ion-Induced Displacive Electric Polarization in FeO$_5$ Bipyramidal Units of Ba,Sr Fe$_{12}$O$_{19}$ Hexaferrites. *Phys Rev B* 2014, *90* (18), 180404. https://doi.org/10.1103/PhysRevB.90.180404.

(71) Bergman, L.; Alexson, D.; Murphy, P. L.; Nemanich, R. J.; Dutta, M.; Stroscio, M. A.; Balkas, C.; Shin, H.; Davis, R. F. Raman Analysis of Phonon Lifetimes in AlN and GaN of Wurtzite Structure. *Phys Rev B* 1999, *59* (20), 12977–12982. https://doi.org/10.1103/PhysRevB.59.12977.

(72) Dokala, R.; Maltoni, P.; Pramanik, P.; Barucca, G.; Varvaro, G.; Edvinsson, T.; Niklasson, G. A.; Peddis, D.; Mathieu, R. High-Temperature Dielectric Properties of Nanostructured SrFe$_{12}$O$_{19}$: Al- Substitution vs Nano-Compositing with CoFe$_2$O$_4$. *In preparation* 2025.

(73) Jahan, N.; Khan, M. N. I.; Hasan, M. R.; Bashar, M. S.; Islam, A.; Alam, M. K.; Hakim, M. A.; Khandaker, J. I. Correlation among the Structural, Electric and Magnetic Properties of Al$^{3+}$ Substituted Ni–Zn–Co Ferrites. *RSC Adv* 2022, *12* (24), 15167–15179. https://doi.org/10.1039/D1RA09354A.

# FIGURES AND TABLES

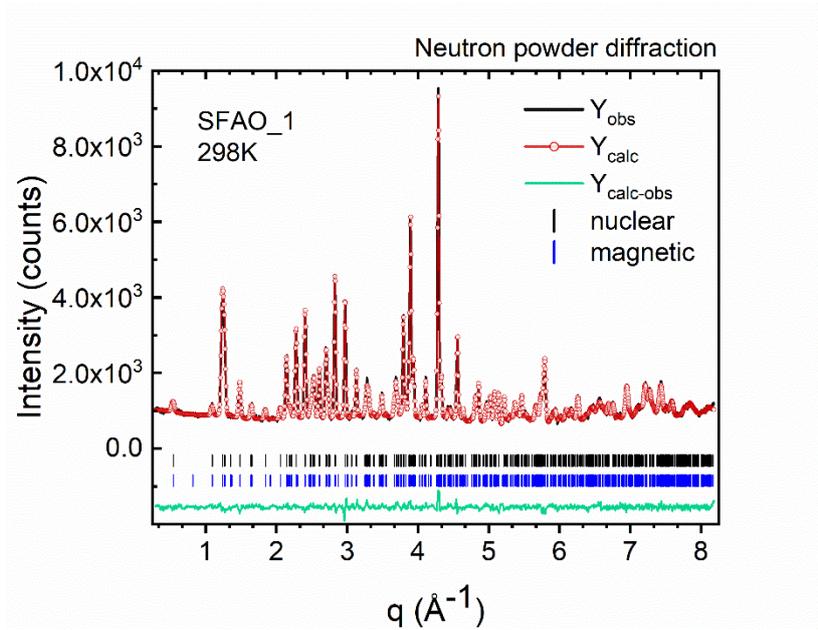

**Figure 1.** Rietveld refinement plot for SFAO_1 sample using neutron powder diffraction data (298K). The ticks identify reflections for the main phases.

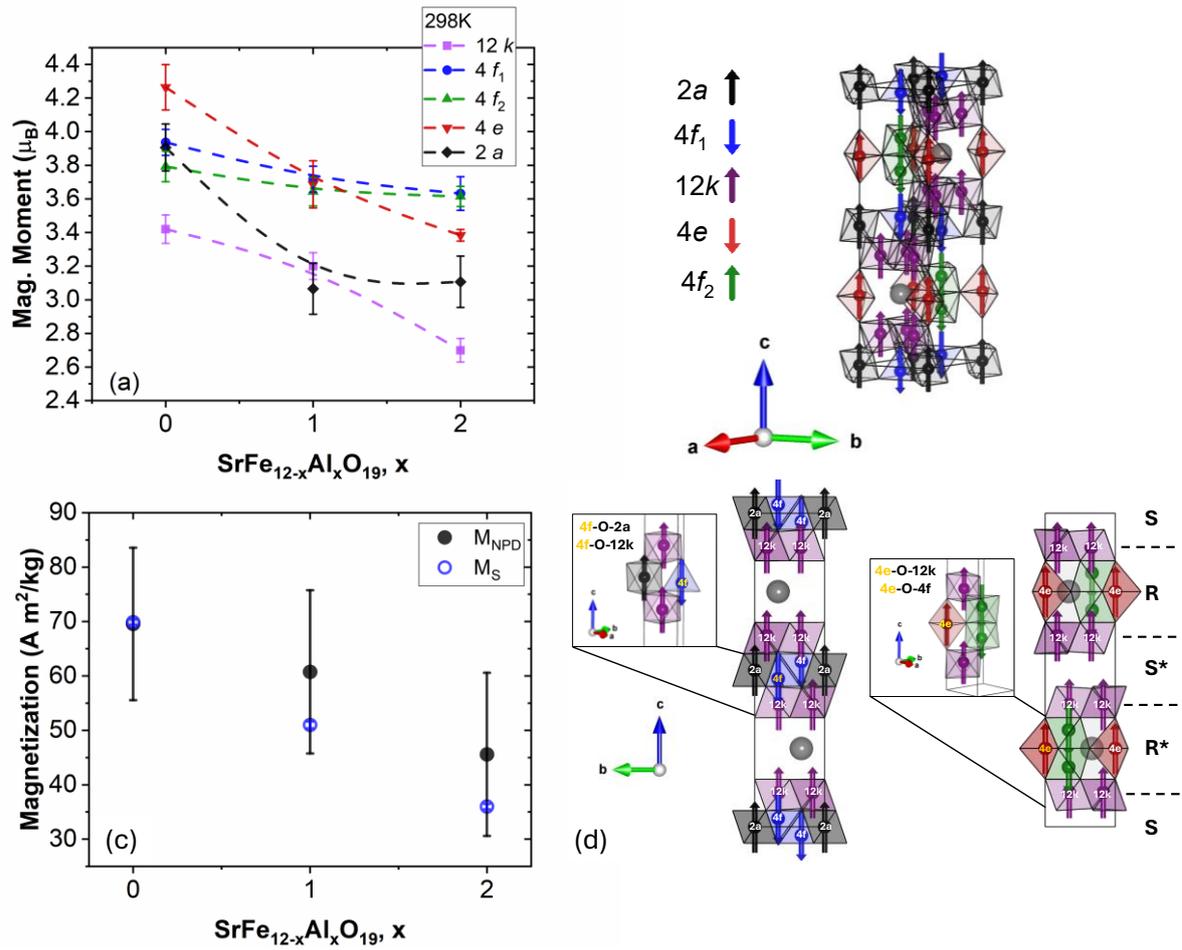

**Figure 2.** a) dependence of the refined magnetic moments of $Fe^{3+}$ at the five different crystallographic sites on aluminium content increasing (298K); (b) schematic representation of the crystal and magnetic structure; (c) measured saturation magnetization ($M_S$), and calculated magnetization from the refined magnetic moments of the NPD data ($M_{NPD}$), as a function of $Al^{3+}$ substitution; (d) schematic representation of the crystal and magnetic structure for tetrahedral 4*f* (left) and bipyramidal 4*e* (right) connections (illustrations made using the software VESTA).

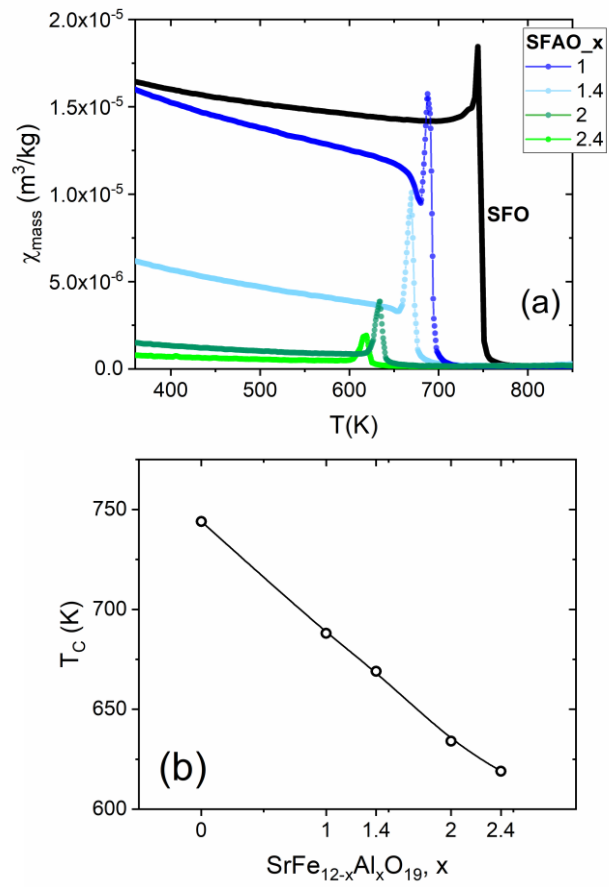

**Figure 3.** (a) magnetic mass susceptibility ($\chi_{mass}$) as a function of increasing temperature; (b) Curie temperatures ($T_C$) estimated from the temperature dependent susceptibility curves.

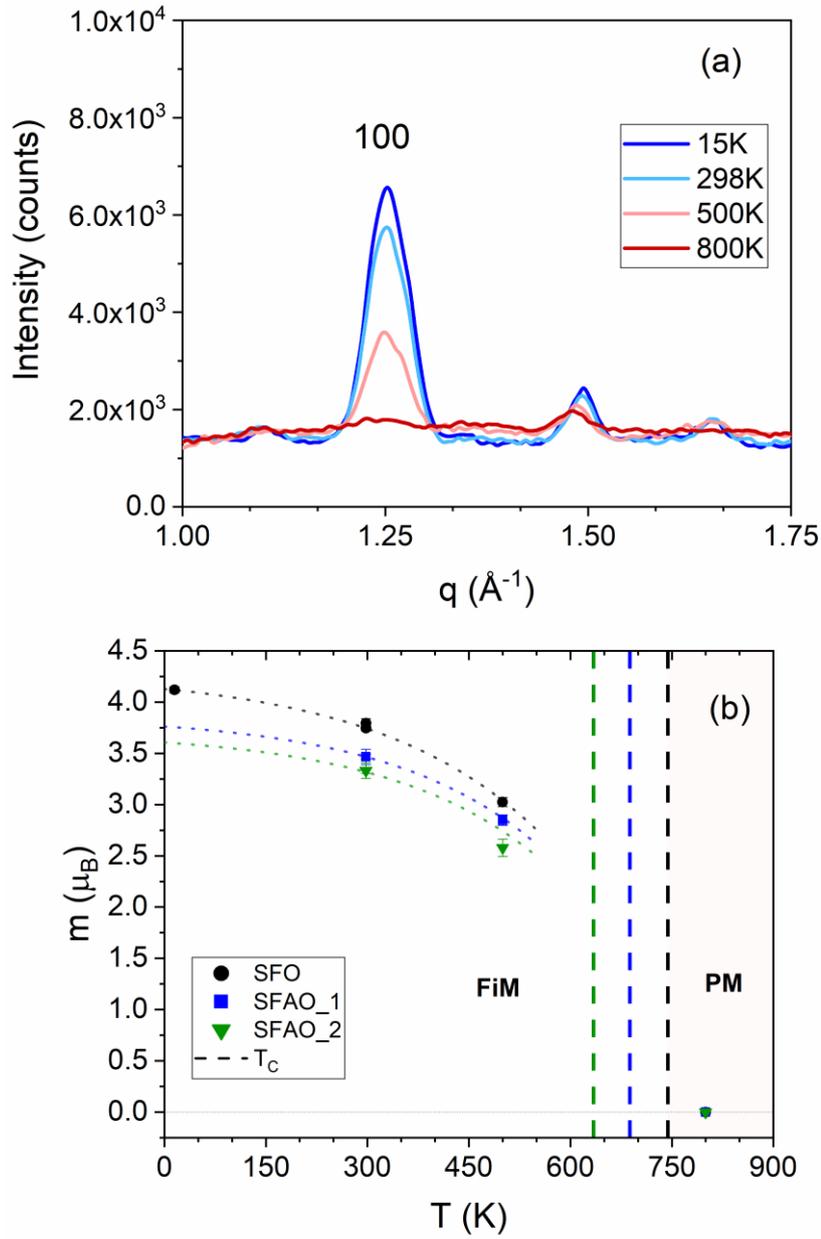

**Figure 4.** (a) detail of the magnetic 100 reflection of NPD patterns for SFO at 15, 298, 500 and 800K; (b) average magnetic moment vs. measurement temperature (dashed lines correspond to corresponding $T_C$ estimated from susceptibility measurements); $m$ values of measurements conducted at 298K after cooling back from 800K are within the error bars of starting values at 298K). Dotted lines are exponential fits as guide to the eye.

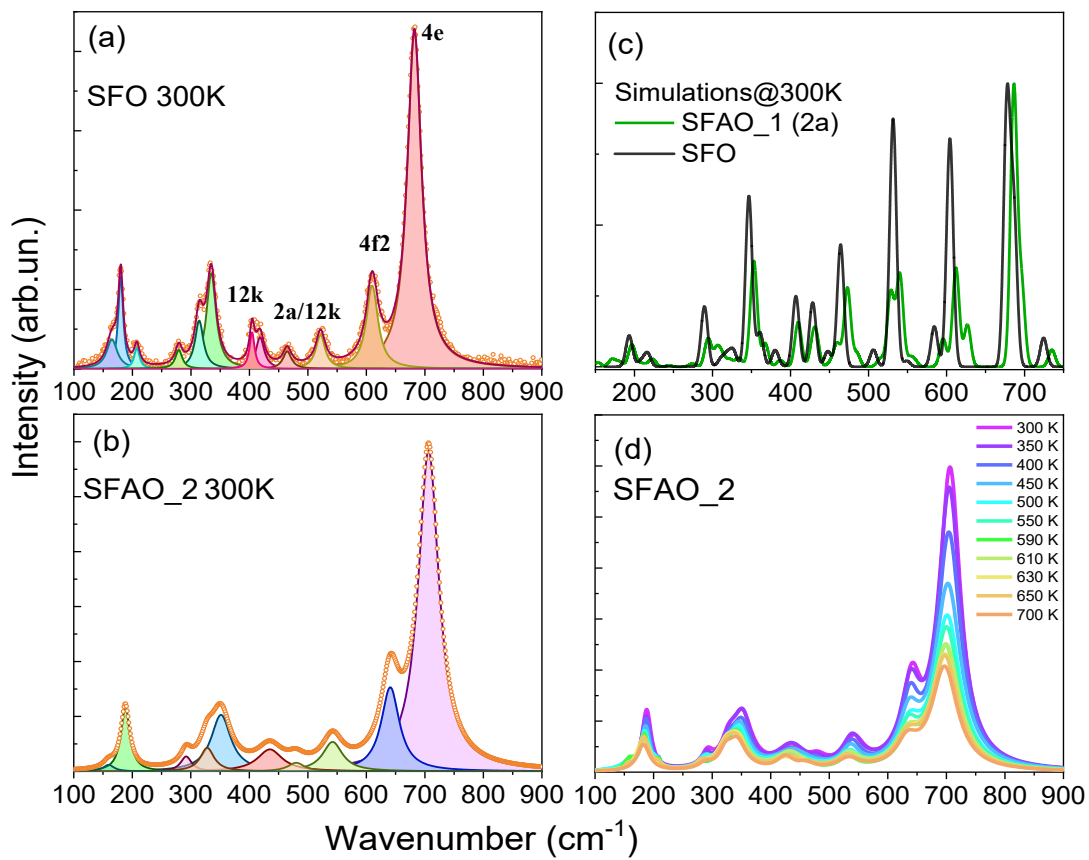

**Figure 5.** (a,b) Raman spectra of SFO and SFAO_2 at 300K, fitted with pseudo-Voigt function; (c) comparison between the calculated Raman spectra of pristine SFO and Al-substituted SFAO_1, where Al occupies the energetically favored 2a site; (d) SFAO_2 Raman spectra recorded at various temperatures.

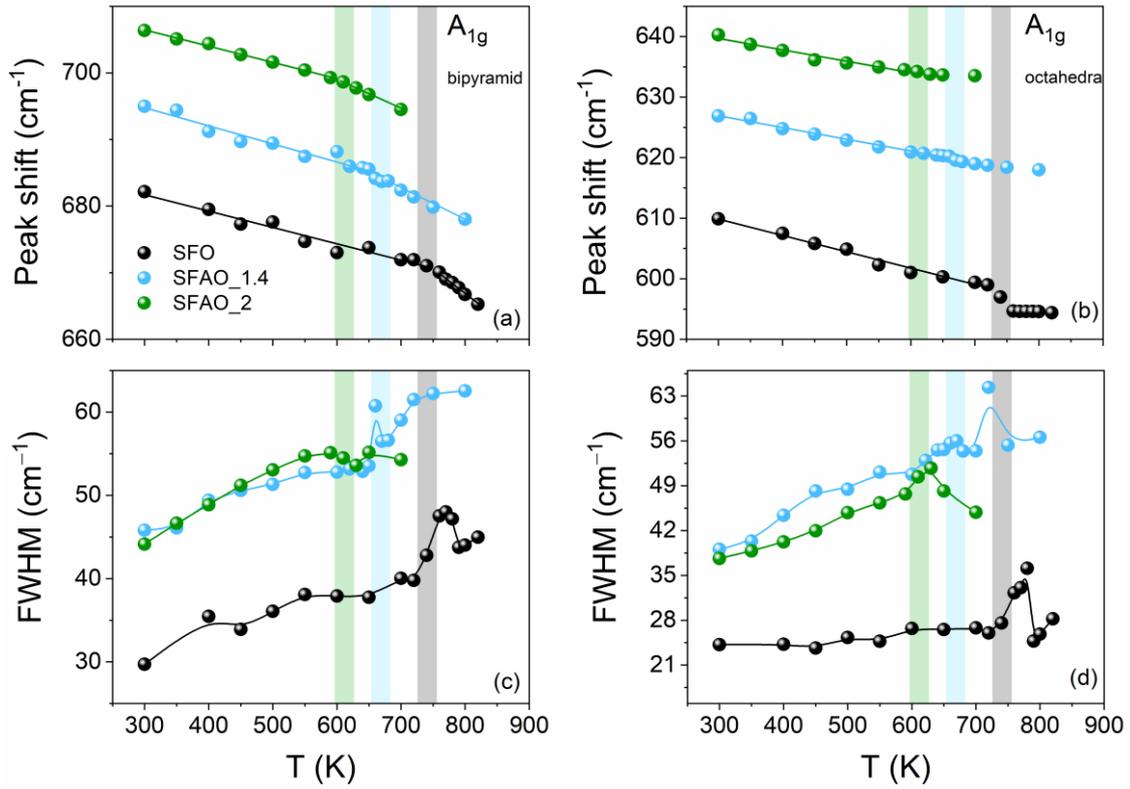

**Figure 6.** Comparisons of temperature dependence of (a,b) peak shift and (c,d) FWHM of some selected Raman-active modes for SFO, SFAO_1.4 and SFAO_2. Peak shift fitted with Balkanski model near the ferrimagnetic transition.

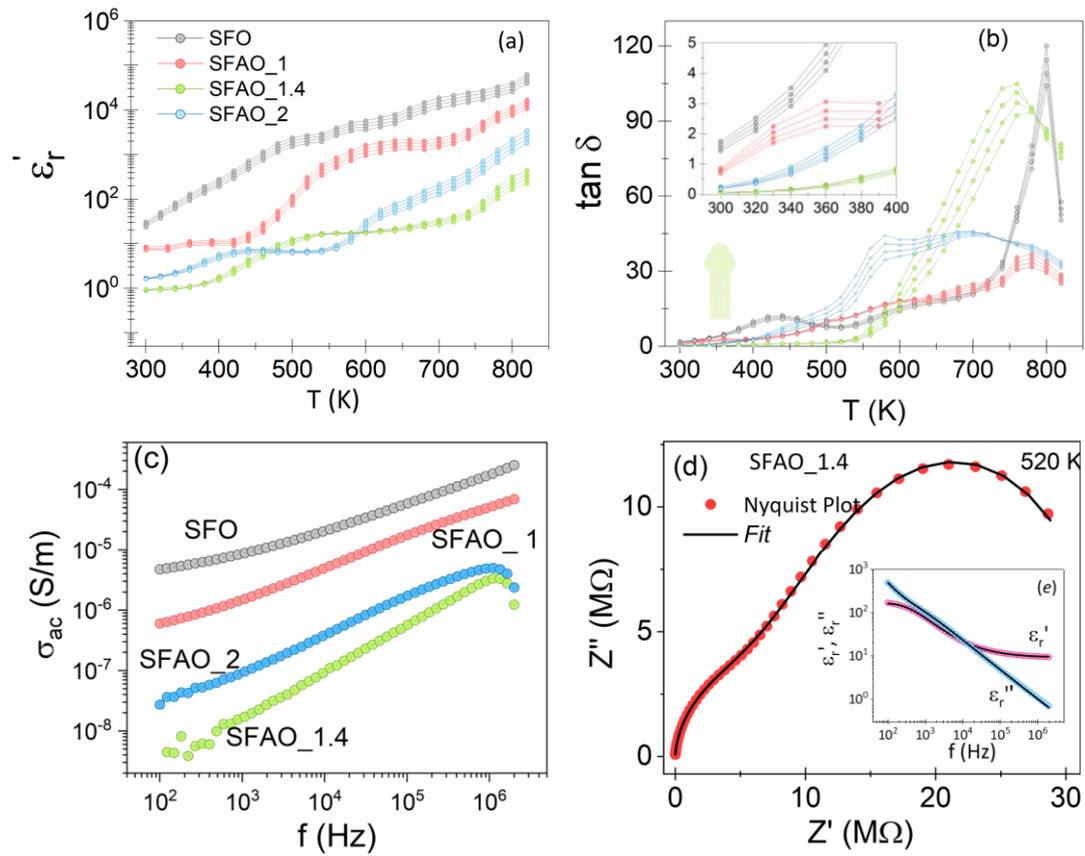

**Figure 7.** High temperature dependence of (a) the real part of the relative dielectric constant, $\varepsilon'_r$ of SFO and SFAO_x ($x$ = 1, 1.4, 2), and (b) their corresponding loss tangent, tan δ in the frequency ranging from 180 Hz to 330 Hz. (c) Frequency dependence of *ac*-conductivity, $\sigma_{ac}$ at room temperature. For the SFAO_1.4 at 520 K, (d) (d) Nyquist plot, and Z″-vs-Z′ fit (main frame) and frequency dependence of the real and imaginary part of the dielectric constant, $\varepsilon_r$ and their corresponding fits (inset, (e)).

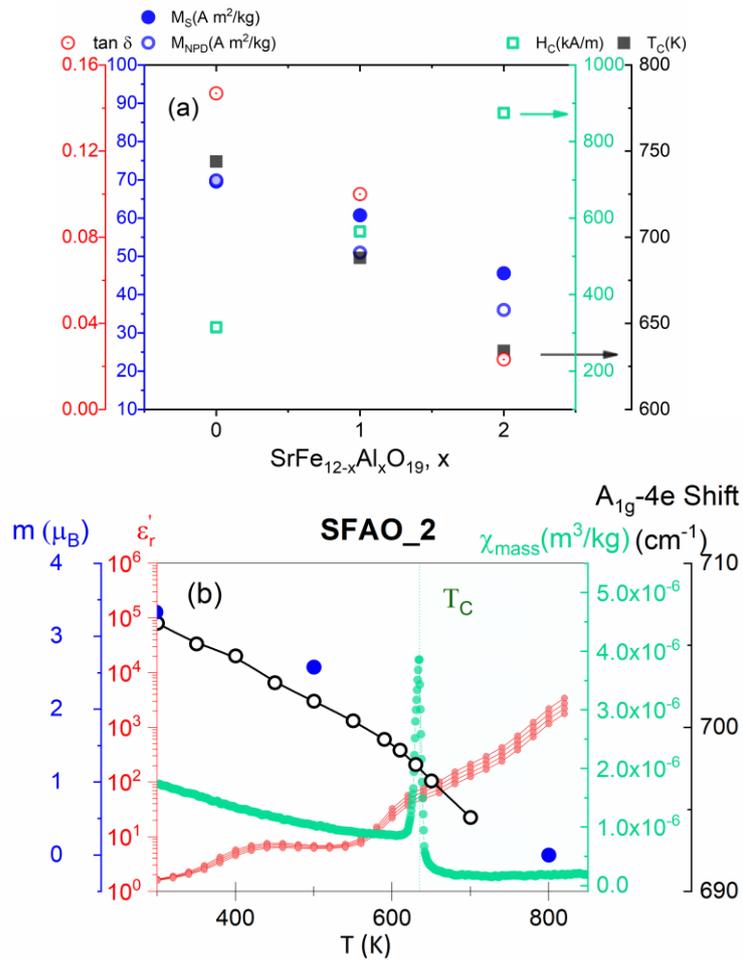

**Figure 8.** (a) summary of estimated magnetic and dielectric parameters ($M_S$, $M_{NPD}$, tan δ, $H_C$, $T_C$) for selected samples, (b) temperature dependent data (m, $\varepsilon'_r$, $\chi_{mass}$, shift of $A_{1g}$-4e) for SFAO_2.